\documentclass[aps,prl,twocolumn,groupedaddress,showpacs]{revtex4}

\usepackage{graphicx}
\usepackage{amsmath}

\begin{document}

\title{Themoelectric Power of High-$T_c$ Cuprate Superconductors Calculated from the Electronic Structure.}

\author{J.G. Storey$^1$, J.L. Tallon$^2$, G.V.M. Williams$^2$}

\affiliation{$^1$Cavendish Laboratory, University of Cambridge, Cambridge CB3 0HE, UK.}

\affiliation{$^2$MacDiarmid Institute, Industrial Research Ltd.,
P.O. Box 31310, Lower Hutt, New Zealand}

\date{\today}

\begin{abstract}
We have calculated the thermopower of the Bi$_2$Sr$_2$CuO$_6$ and Bi$_2$Sr$_2$CaCu$_2$O$_8$ superconductors using an ARPES-derived dispersion, with a model pseudogap, and a marginal-Fermi liquid scattering rate that has a minimum with respect to energy at the van Hove singularity (vHs). Good fits with data are achieved across the entire phase diagram, thus confirming the dispersions, the locations of the vHs and the dominance of the diffusion thermopower over the phonon drag contribution.
\end{abstract}


\maketitle

The thermoelectric power (TEP) of the high-temperature cuprate superconductors was early shown\cite{OBERTELLI} to exhibit universal behaviour. Roughly speaking, the linear part of the TEP takes the form $S(T)=S_0-\alpha T$. $S_0$ is positive and large in the underdoped regime and decreases with doping, becoming negative in the overdoped regime. This behaviour has been modelled in terms of separate metallic diffusion and phonon drag contributions by Trodahl\cite{PHONONDRAGMODEL}. In the model, the TEP comprises the sum of a negative linear metallic component, and a positive phonon drag component which rises at low temperatures before saturating at high temperatures.

\begin{figure}
\centering
\includegraphics[width=50mm,clip=true,trim=0 0 0 0]{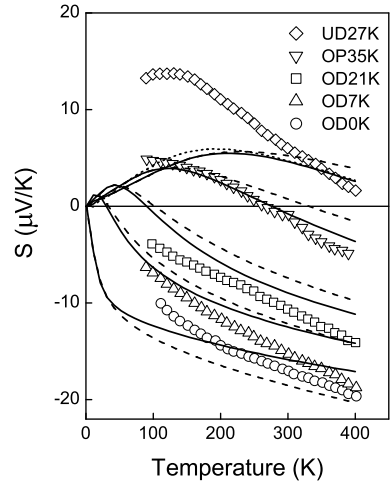}
\caption{Measured and calculated thermopower of Bi-2201\cite{KONDOTEP}. The curves are calculated assuming a constant mean free path. Dashed curves include corrections to the chemical potential for temperature. The dotted curve shows the effect of a pseudogap in the density of states. The values of $E_F-E_{vHs}$ from underdoped to overdoped are 70, 42, 16, 6 and -4meV.}
\label{KONDOTEPFIG}
\end{figure}

Alternatively, McIntosh and Kaiser\cite{MCINTOSH} calculated the TEP from a model electronic structure containing a van Hove singularity (vHs), and were able to explain the TEP in terms of the diffusion component alone. Here the reduction in the positive part of the TEP with doping results from the approach to, and traversal of, the vHs by the Fermi level ($E_F$). Experimental support for this picture has been published by Kondo \textit{et al.}\cite{KONDOTEP}. They measured both the electronic structure, by angle-resolved photoemission spectroscopy (ARPES), and the TEP of a series of Bi-2201 samples. The TEP was calculated from the electronic structure and a good correspondence with the data was obtained for over- and optimally doped samples. Their data and calculations (replicated by the present authors) are shown in Fig.~\ref{KONDOTEPFIG}. The calculations show the positive peak in the TEP reducing and disappearing as $E_F$ approaches and traverses the vHs.

The calculations by Kondo \textit{et al.}, which have no free parameters but assume a constant mean free path and a $T$-independent chemical potential ($\mu$), are unable to reproduce the magnitude of the TEP in the underdoped regime. This is perhaps not surprising given that the pseudogap, a key feature of the underdoped regime, is not taken into account. Here we extend their approach by including models for the pseudogap and the scattering rate.

The diffusion thermopower is given by\cite{ALLEN}
\begin{equation}
S(T)=\frac{1}{\left|e\right|T}\frac{\int_{-\infty}^{\infty}{\sigma\left(\epsilon\right)\left(\epsilon-\mu\right)\left(\frac{\partial{f\left(\epsilon\right)}}{\partial{\epsilon}}\right)d\epsilon}}{\int_{-\infty}^{\infty}{\sigma\left(\epsilon\right)\left(-\frac{\partial{f\left(\epsilon\right)}}{\partial{\epsilon}}\right)d\epsilon}}
\label{TEPEQ}
\end{equation}
$\sigma(\epsilon)$ is the spectral conductivity, which under the Boltzmann formalism is given by
\begin{equation}
\sigma(\epsilon)=\frac{e^2}{V}\sum_{\textbf{k}}{v_{\alpha}(\epsilon,\textbf{k})\ell_{\alpha}(\epsilon,\textbf{k},T)\delta[\epsilon(\textbf{k})-\epsilon]}
\label{SIGMAEQ}
\end{equation}
$\ell_{\alpha}$ is the mean free path given by
\begin{equation}
\ell_{\alpha}=v_{\alpha}(\epsilon,\textbf{k})\cdot\tau(\epsilon,\textbf{k},T)
\label{MFPEQ}
\end{equation}
where $v_{\alpha}=\partial{\epsilon(\textbf{k})}/\partial{k_{\alpha}}$ is the group velocity of the conduction electrons, $\tau(\epsilon,\textbf{k},T)$ is the relaxation time and $\alpha=x$ or $y$.

\begin{figure}
\centering
\includegraphics[width=70mm,clip=true,trim=0 0 0 0]{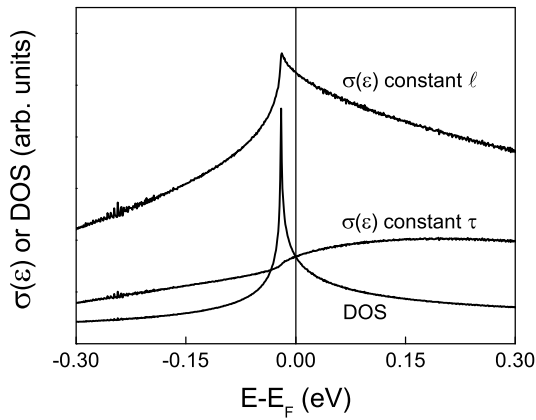}
\caption{The density of states of Bi-2201 calculated from the ARPES-derived energy-momentum dispersion and the conductivity calculated from Eqn.~\ref{SIGMAEQ} assuming a constant mean free path and constant relaxation time.}
\label{KONDODOSFIG}
\end{figure}

The TEP as defined by Eq.~\ref{TEPEQ} is a measure of the asymmetry in $\sigma(\epsilon)$ about $\mu$ via the spectral window $(\epsilon-\mu)[\partial{f(\epsilon)}/\partial{\epsilon}]$. The saddle-points in the measured energy-momentum dispersion $\epsilon(\textbf{k})$\cite{KONDOTEP} give rise to a vHs in the density of states as shown in Fig.~\ref{KONDODOSFIG}. Assuming a constant mean free path, $\sigma(\epsilon)\propto\sum_{\textbf{k}}{v_{\alpha}\delta[\epsilon(\textbf{k})-\epsilon]}$ is just a velocity-weighted density of states. Because the velocities near the saddle points are small, the peak in $\sigma(\epsilon)$ is less prominent than the vHs is in the DOS.

As a first step we have corrected $\mu$ so that the carrier concentration 
\begin{equation}
n=2\int{f(\epsilon)N(\epsilon)d\epsilon}
\label{nEQ}
\end{equation}
remains constant with temperature resulting in the dashed curves shown in Fig.~\ref{KONDOTEPFIG}. 
This affects the magnitude but not the general behaviour of the thermopower. $\mu$ is corrected for temperature in all subsequent calculations.

Next we include a model for the pseudogap that we have used previously to describe the electronic entropy\cite{STOREYENTROPY,STOREYPG}, superfluid density\cite{STOREYENTROPY} and Raman response\cite{STOREYRAMAN,STOREYRAMAN2}. It is based on ARPES results and takes the form\cite{STOREYENTROPY} 
\begin{equation}
E_{g}=\left\{
\begin{array}{ll}
E_{g,max}\cos{\left(\frac{2\pi\theta}{4\theta_{0}}\right)} & (\theta<\theta_{0})\\
\\
E_{g,max}\cos{\left(\frac{2\pi(\theta-\pi/2)}{4\theta_{0}}\right)} & (\theta>\frac{\pi}{2}-\theta_{0})\\
\\
0 & \mbox{otherwise}
\end{array}\right.
\label{PGEQ}
\end{equation}
where $0\leq\theta\leq\pi/2$ is the angle subtended at $(\pi,\pi)$ by the points $(k_{x},k_{y})$ and $(\pi,0)$. 
$\theta_{0}$ reproduces the Fermi arc phenomenology and is given by\cite{STOREYPG}
\begin{equation}
\theta_{0}=\theta_{0,max}\left[1-\tanh{\left(\frac{T}{2T^{*}}\right)}\right]
\label{THETA0EQ}
\end{equation}
where $T^{*}=E_{g,max}/k_{B}$. The non-states-conserving nature of the pseudogap is implemented by removing states with $|\epsilon(\textbf{k})-\mu|<E_g$ from Eqs.~\ref{TEPEQ}, \ref{SIGMAEQ} and \ref{nEQ}. This is equivalent to such states having very short lifetimes.

The dotted curve in Fig.~\ref{KONDOTEPFIG} shows the TEP calculated for the most underdoped sample assuming a constant mean free path and including a pseudogap with $E_{g,max}=18$meV and $\theta_{0,max}=33^\circ$. There is a boost in the TEP below 250K and an increase in slope above 250K but the effect is too small to significantly improve the fit. Under the constant $\ell$ assumption the model pseudogap does not introduce enough additional asymmetry in $\sigma(\epsilon)$ about $\mu$. This brings us to the scattering rate.
\begin{figure}
\centering
\includegraphics[width=70mm,clip=true,trim=0 0 0 0]{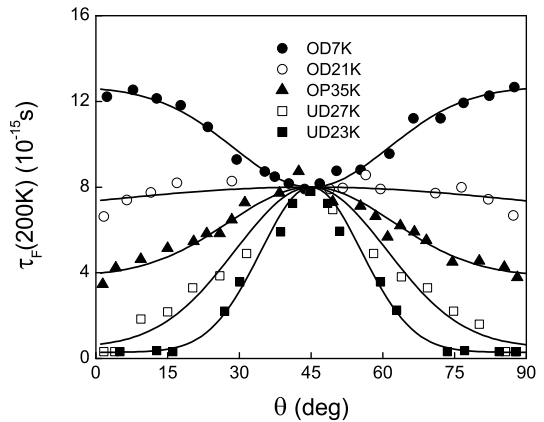}
\caption{Angle dependence of the lifetime of Bloch states at $E_F$ determined by ARPES measurements at 200K\cite{KONDOTAU}. The OD7K data indicates long lifetimes near the saddle points at 0 and 90$^\circ$.}
\label{ARPESTAUFIG}
\end{figure}

We adopt a scattering rate of the form
\begin{equation}
\hbar\tau^{-1}\left(\epsilon,T\right)=\lambda\sqrt{\left(\pi{}k_BT\right)^2+\left(\epsilon-E_{vHs}\right)^2}+a
\label{TAUEQ}
\end{equation}
where $\lambda$ is a coupling constant. $a$ takes a fixed value of 1meV and is included to prevent $\tau$ from becoming infinite.
The square root term in Eq.~\ref{TAUEQ} is similar to the implementation of the max($|\omega|,T$) marginal Fermi liquid single-particle scattering rate employed by Abrahams and Varma\cite{ABRAHAMS}, but it differs in the location of the minimum with respect to energy.  In order to produce a peak in $\sigma(\epsilon)$ at the energy of the vHs, Eq.~\ref{TAUEQ} must have a minimum at $E_{vHs}$  rather than at $E_F$. This is because $\sum_{\textbf{k}}{v_{\alpha}^2\delta[\epsilon(\textbf{k})-\epsilon]}$ does not exhibit a peak at $E_{vHs}$, as shown by the plot of $\sigma(\epsilon)$ for constant $\tau$ in Fig.~\ref{KONDODOSFIG}. Supporting evidence for a small scattering rate at the vHs comes from angle dependent ARPES measurements of the lifetime of Bloch states at $E_F$\cite{KONDOTAU}. The data, reproduced in Fig.~\ref{ARPESTAUFIG}, shows the
average lifetime around the Fermi surface increasing with doping.
Furthermore, for the most overdoped sample (filled circles) the relaxation time
is maximal at 0 and 90 degrees. Because $E_F$ lies close to the vHs at
this doping this indicates that the scattering rate is small near the saddle
points. Similar behaviour is also found in overdoped Tl-2201\cite{PEETS}. The $\epsilon-E_{vHs}$ dependence arises from a more complex underlying $k$-dependent relaxation time and importantly introduces asymmetry in $\sigma(\epsilon)$ about $\mu$. This in turn gives rise to asymmetry in the pseudogap, which increases the TEP.

\begin{figure}
\centering
\includegraphics[width=70mm,clip=true,trim=0 0 0 0]{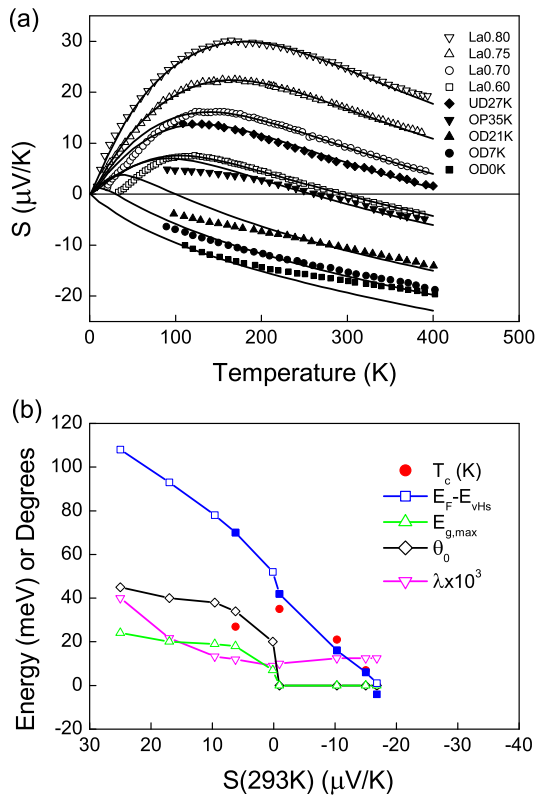}
\caption{(a) Fits to Bi-2201 TEP data of Refs.~\cite{KONDOTEP} (closed symbols) and \cite{OKADA} (open symbols) using models for the pseudogap and scattering rate given by Eqs.~\ref{PGEQ}, \ref{THETA0EQ} and \ref{TAUEQ}. (b) Parameters extracted from the fits in (a). Solid symbols are experimentally determined.}
\label{2201FITFIG}
\end{figure}

Fits to the Bi-2201 data of Kondo \textit{et al.} and Okada \textit{et al.}\cite{OKADA} are shown in Fig.~\ref{2201FITFIG}(a). The parameters used in the fits are plotted in Fig.~\ref{2201FITFIG}(b) as a function of TEP at 293K. We stress here that $\lambda$ is the only truly free parameter used in the model. $E_F-E_{vHs}$, $E_{g,max}$ and $\theta_{0,max}$ can be determined by experiment. Experimental values for $E_F-E_{vHs}$ are shown by the solid blue squares. For the most overdoped dataset we find that a value of 1meV for $E_F-E_{vHs}$ gives a better fit than the -4meV measured experimentally. We note that the ARPES measurements were performed at 200K. The calculated chemical potential corrections support a drift of a few meV over this temperature range due to the close proximity of the vHs.

\begin{figure}
\centering
\includegraphics[width=70mm,clip=true,trim=0 0 0 0]{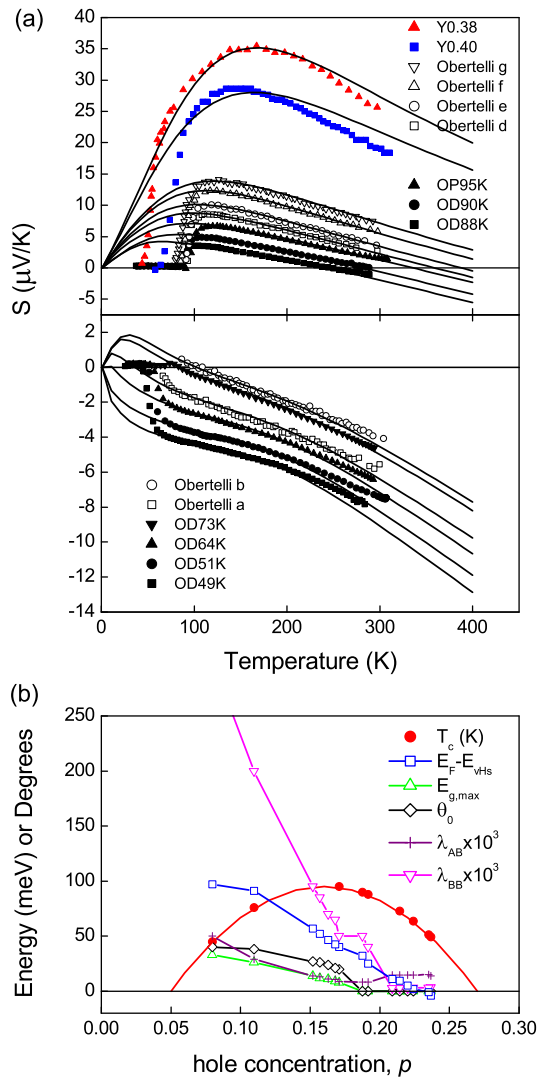}
\caption{(a) Fits to our own Bi-2212 TEP data (black solid symbols) as well as data from Refs.~\cite{OBERTELLI,MANDRUS,MUNAKATA}. (b) Parameters extracted from the fits in (a).}
\label{2212FITFIG}
\end{figure}

We have also performed fits to Bi-2212 TEP data using an ARPES-derived bilayer dispersion\cite{STOREYPHASE}. Separate instances of Eq.~\ref{TAUEQ} were applied to the antibonding and bonding bands with coupling constants $\lambda_{AB}$ and $\lambda_{BB}$  respectively. Recently we used the same dispersion and pseudogap model to fit electronic entropy data\cite{STOREYPG}. From those fits we were able to extract values for $E_F-E_{vHs}$, $E_{g,max}$ and $\theta_{0,max}$ that closely matched values determined directly from ARPES. Here we have used those values as a guide for fitting the TEP data, leaving only $\lambda_{AB}$ and $\lambda_{BB}$ as truly free parameters.
Fits to our own Bi-2212 data as well as data from Obertelli \textit{et al.}\cite{OBERTELLI}, Mandrus \textit{et al.}\cite{MANDRUS} and Munakata \textit{et al.}\cite{MUNAKATA} are shown in Fig.~\ref{2212FITFIG}(a). The corresponding parameters as a function of hole concentration are shown in Fig.~\ref{2212FITFIG}(b). 

From the fits, the curvature in the overdoped data above 100K is seen to arise from the bonding band vHs, which is located approximately 100meV below the antibonding band vHs. The same conclusion was reached by Takeuchi \textit{et al.}\cite{TAKEUCHI} who performed calculations using a bilayer dispersion and a constant mean free path.

\begin{figure}
\centering
\includegraphics[width=70mm,clip=true,trim=0 0 0 0]{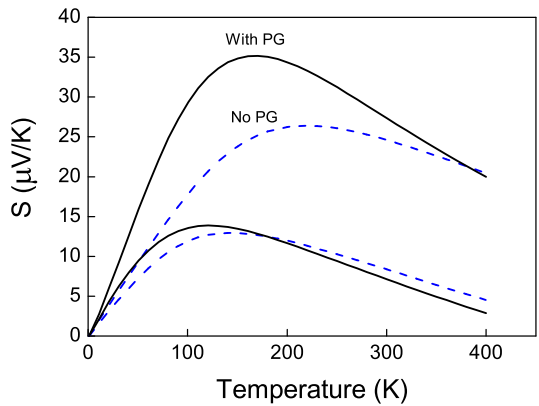}
\caption{Two of the curves from Fig.~\ref{2212FITFIG}(a) recalculated with $E_{g,max}$ set to zero (dashed curves) to illustrate the effect of the pseudogap on the TEP.}
\label{NOPGFIG}
\end{figure}

The effect of the pseudogap on the TEP is illustrated in Fig.~\ref{NOPGFIG} for two different dopings, by setting $E_{g,max}$ to zero while keeping $E_F-E_{vHs}$ and $\lambda$ unchanged. The pseudogap boosts the TEP at low temperatures and increases the high temperature slope. The peak in the TEP results from the combined effects of the pseudogap and the vHs, not just the pseudogap alone. Attempts\cite{TAKEMURA} to determine the pseudogap temperature $T^*$ by a simple scaling analysis are therefore somewhat naive. The relevance of the band structure to the scaling behaviour of the TEP is also discussed by Okada \textit{et al.}\cite{OKADA}.

The coupling constants are roughly constant in the overdoped regime and increase systematically with the opening of the pseudogap. The behaviour is reminiscent of the isotope effect coefficient\cite{PRINGLE}.

\begin{figure}
\centering
\includegraphics[width=70mm,clip=true,trim=0 0 0 0]{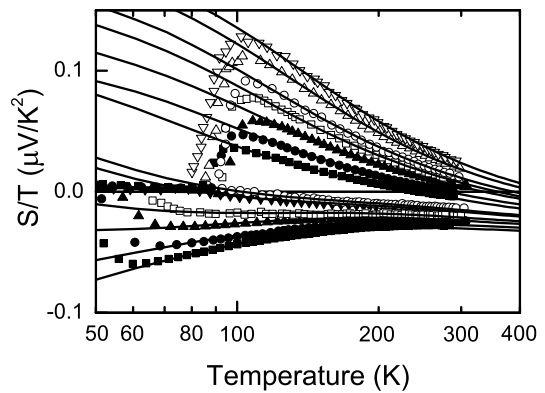}
\caption{$S/T$ versus $\log(T)$ for the data in Fig.~\ref{2212FITFIG}(a). Symbols have the same meaning. The fanning out behavior (which near $T=0$ cannot be scaled) is due to crossing the vHs and not a quantum critical point (where scaling would apply near $T=0$).}
\label{SCALINGFIG}
\end{figure}

Recently Daou \textit{et al}.\cite{DAOU}, have measured the TEP of La$_{1.6-x}$Nd$_{0.4}$Sr$_x$CuO$_4$ and for $x=0.24$ found a linear behavior when $S/T$ is plotted versus $\log(T)$. Below this doping the low-$T$ data (for $x=0.2$) fanned upwards while above this doping, for $x=0.3$, it fanned downwards. (In the latter case they used data for the Nd-free compound). They deduced from this the presence of a quantum critical point at $x=0.24$, which in their view locates the termination of the pseudogap line $T^*(x)$. We feel this conclusion is not warranted. Leaving aside the paucity of doping points, the behavior is just what is expected from Figs.~\ref{2201FITFIG}  and \ref{2212FITFIG} for traversing a vHs. In Fig.~\ref{SCALINGFIG} we replot the latter data as $S/T$ versus $\log(T)$. It shows a distinctive fanning out at low $T$ either side of the vHs which is located between the 3rd and 4th to bottom curves. It should also be noted that La$_{2-x}$Sr$_x$CuO$_4$ exhibits a peak in the DOS and in the static susceptibility precisely at $x=0.24$ indicating the location of the vHs crossing\cite{ENTROPYDATA2}.

In summary, we have calculated the TEP of high-$T_c$ cuprate superconductors using an ARPES-derived dispersion, with a model pseudogap, and a marginal-Fermi-liquid scattering rate that has a minimum with respect to energy at the vHs. Good fits with data are achieved across the phase diagram confirming the overall ARPES-derived dispersion and the location of the vHs. Our results show that the diffusion thermopower dominates over the phonon drag contribution.

\end{document}